\documentclass[usenatbib]{mn2e}
\usepackage[totalwidth=480pt, totalheight=680pt]{geometry}

\newcommand{\angstrom}{\ang}
\newcommand{\etal}{\mbox{et\ al.\ }}
\newcommand{\ang}{\,\mbox{\AA}}

\newcommand{\ergsec}{\,\mbox{$\mbox{erg}\,\mbox{s}^{-1}$}}

\newcommand{\msun}{\,\mbox{$\mbox{M}_{\odot}$}}

\title[Constraints on AGN accretion disc viscosity]
{Constraints on AGN accretion disc viscosity derived from continuum
variability}
\author[R.L.C. Starling, A. Siemiginowska, P. Uttley \& R. Soria]
{Rhaana L. C. Starling$^{1,2}$\thanks{E-mail: rlcs@mssl.ucl.ac.uk}, Aneta Siemiginowska$^2$, Phil Uttley$^3$
~$\&$~ Roberto Soria$^1$  \\
$^1$ Mullard Space Science Laboratory, University College London, Holmbury St.
Mary, Dorking, Surrey RH5 6NT, UK \\
$^2$ Harvard-Smithsonian Center for Astrophysics, 
60 Garden Street, Cambridge, MA 02138, USA \\
$^3$ Dept. of Physics and Astronomy, University of Southampton, Hampshire SO17 1BJ, UK}

\begin{document}
\date{Accepted 4 September 2003. Received ; in original form 14 July 2003}

\pagerange{\pageref{firstpage}--\pageref{lastpage}} \pubyear{2003}

\maketitle

\label{firstpage}

\begin{abstract} 
We estimate a value of the viscosity parameter in AGN accretion discs for the PG
quasar sample. We assume that optical variability on time-scales of
months to years is caused by local instabilities in the
inner accretion disc. Comparing the observed variability time-scales
to the thermal time-scales of $\alpha$-disc
models we obtain constraints on the viscosity parameter for the
sample. We find that, at a given $L/L_{\rm Edd}$, the entire sample is consistent with a single
value of the viscosity parameter, $\alpha$. We obtain constraints
of $0.01 \le \alpha \le 0.03$ for $0.01 \le L/L_{\rm Edd} \le 1.0$.
This narrow range suggests that these AGN are all seen in a single state, with
a correspondingly narrow spread of black hole masses or accretion rates. The
value of $\alpha$ we derive is consistent with predictions by current simulations in
which MHD turbulence is the primary viscosity mechanism. 
\end{abstract}


\section{Introduction}
The transport of angular momentum remains a major unknown in accretion disc theory. Shakura \& Sunyaev (1973) suggested that
magnetic fields and turbulent motions are likely candidates for the
angular momentum transport. They argued that turbulence is a dominant
factor and postulated that the stress is proportional to the
local pressure. The proportionality parameter $\alpha$, which parametrizes the underlying transport process, describes the
efficiency of the turbulent transport. The $\alpha$-parameter also
represents the scale of the largest possible turbulent cell. The
$\alpha$ parameterization provided a base for developing the accretion
disc theory and linked the theory with the observations.
                                                   
More recent work has shown that in weakly magnetized discs MHD turbulence
can provide the majority of the outward transport of
angular momentum \citep{BH91}.
Balbus \& Papaloizou (1999) reviewed the $\alpha$-disc models and found
that MHD turbulence follows the $\alpha$ prescription very
closely. They show that the vertically averaged disc with
viscosity due to magnetorotational instabilities is described by a
similar set of radial equations to the $\alpha$-discs. In particular
the local energy dissipation rate is determined by the global disc
parameters as it is in the $\alpha$-discs. 
However, global 3-D numerical
simulations of accretion discs are extremely complex \citep{MHD} and so most work to date has
been done using local 2-D shearing box, cylindrical and axisymmetric limit
models. All these methods have provided predictions of the $\alpha$-parameter
ranging from $\sim$0.005 \citep{Branden} to $\sim$0.6 \citep{HGB95}.

In order to determine the magnitude of the viscosity
better estimates from larger samples of all categories of accreting
objects are needed with independent methods for estimating $\alpha$.
Since the spectrum of a disc at any particular instant is not strongly
dependent on its viscosity the most effective way to estimate the
viscosity is through the variability of the disc luminosity. A number of studies of
this kind have been done for accreting binary systems since the
time-scales of variability are optimal for observation, generally
ranging from days down to a fraction of a second. 

In the case of AGN we observe typical optical fluctuation time-scales
of weeks to years although there have been only a few long-term monitoring
programs to provide such data (eg. Pica et
al. 1988; Giveon et al. 1999, hereafter G99). 
At optical and ultraviolet frequencies the majority of the continuum
luminosity seen in AGN is held to be emission from the accretion disc
\citep{MS} and large variability has been observed in
these wavebands for many such objects \citep{Webb}. The physical origin of optical flux variations has not been
pinned down, but if disc emission dominates the spectrum at optical
wavelengths the variability is expected to be related to the disc
instabilities.

In this study we use a subset of the Palomar Green sample of AGN (G99), spanning approximately seven years of data, to derive
limits on the viscosity parameter for the the sample as a whole via
the method described in Siemiginowska \& Czerny (1989, hereafter
SC89). We assume any fluctuations in the lightcurves are due to an
instability arising in the innermost disc regions and can be
associated with the thermal time-scale. We do not identify the exact
physical process responsible for this instability. It can be related
to the Lightman-Eardley instabilities in the radiation pressure
dominated region of the disc \citep{LE}, but our method does not require
modelling of the instability and therefore could be applied to any
type of instability which may be present in the inner disc. Using a
set of multi-temperature blackbody disc models, based on the standard
$\alpha$-disc as prescribed by Shakura \& Sunyaev (1973), we
calculate the contribution to the optical spectrum from the outer,
stable disc and assume that there is no contribution from the
innermost parts of the disc where the instability has
developed. Comparing the variability time-scales of the sources with
the thermal time-scales of the models allows us to constrain the
$\alpha$-parameter for the sample.

\section{Data Sample}
The data are taken from the optically selected Palomar-Green (PG; Schmidt \&
Green 1983) sample. A subset of 42 bright ($-28<M_{B}<-22$) and nearby AGN
$(0.06<z<0.37)$ in the northern hemisphere were observed at the Wise
Observatory, Israel (G99) in 1991-1998. Observations were made with a CCD camera mounted on the 1m
optical telescope and span 7 years with a typical sampling interval of
39 days - a higher, more even sampling interval than that of previous
programs (eg. Pica et al. 1988). The wavelength range covers the Johnson \emph{B} band, 3600-5600$\angstrom$
($\lambda_{\rm central}$ = 4400$\angstrom$), which is a region of the
spectrum believed to be dominated by the disc emission. The PG sample is
statistically complete with good photometric accuracy over this
wavelength range ($\sim$0.02 mag. at 4400$\angstrom$). G99 assume a
Hubble constant $H_{0}$ = 70 kms$^{-1}$Mpc$^{-1}$, a deceleration
parameter of $q_{0}$ = 0.2 and a power-law shaped continuum,
$F_{\nu}\propto\nu^{-\alpha}$ where the spectral index $\alpha$ = 0.5
for the K-correction. This is a flux-limited sample so we expect to
see selection effects as discussed in Schmidt \& Green (1983). This sample has also been studied at a number of other wavelengths
including \emph{R} band (G99), radio \citep{Kell}, IR
\citep{Neu} and X-ray \citep{Lao} and its optical polarization properties were
studied by Berriman et al. (1990). Measurements of the masses of 17 PG quasars
included in this subsample have been made by reverberation mapping \citep{Kasp}.

From the lightcurves we derived a two-folding time-scale as a
characteristic time-scale for each source. The two-folding time-scale gives the time for an objects' luminosity to
change by a factor of two (either an increase or a decrease) which can
be used as a measure of variability in both the data and the models
(O'Brien, Gondhalekar \& Wilson 1988 and references therein; SC89).
\begin{equation}
        \tau_{\rm 2} = \tau_{\rm s} \frac{f_{\rm min}}{(f_{\rm max}-f_{\rm min})}   
\end{equation}
where $f_{\rm max}$ and $f_{\rm min}$ are the maximum and minimum observed flux levels
respectively, $\tau_{\rm s}$ = $\tau_{\rm max}-\tau_{\rm min}$ is the time between maximum and
minimum flux levels and $\tau_{\rm 2}$ is the two-folding time-scale. The lightcurves for this data set do not show flux variations of a
factor of two, in fact $\sigma_{B}$/$\emph{B}\le$34$\%$, in which case
$\tau_{\rm 2}$ is the time over which a linearly extrapolated flux variation would
increase or decrease the observed flux by a factor of two. These timescales
were then corrected for the effects of cosmological time dilation.

In order to calculate
$f_{\rm min}/(f_{\rm max}-f_{\rm min})$ we used the standard
transformation from magnitude to flux,
\begin{equation}
        m_{\rm min} - m_{\rm max} = -2.5\log \left(\frac{f_{\rm min}}{f_{\rm max}}\right)
\end{equation}
The luminosity is calculated using the following
three equations relating apparent magnitude, flux and luminosity at 4400$\ang$
(eg. Weedman 1998),
\begin{equation}
\log \frac{F_{\lambda,4400\angstrom}}{\rm erg~cm^{-2}~s^{-1}~\ang^{-1}} = -0.4(m_{B}+20.42)
\end{equation}
\begin{equation}
  L_{\lambda,4400\ang} = 4 \pi d^{2}(1+z)^{2} F_{\lambda,4400\angstrom}
\end{equation}
\begin{equation}
    d = \frac{c[q_{0}z+(q_{0}-1)(\sqrt{1+2q_{0}z}-1)]}{q_{0}^{2}
    H_{0}(1+z)}
\end{equation}
Here, $F_{\lambda,4400\ang}$ and $L_{\lambda,4400\ang}$ are the median flux
and luminosity at 4400$\ang$, $c$ is the speed of light in a vacuum. 
The apparent magnitudes, $m_{B}$, are taken from G99 and corrected for the
line-of-sight extinction, with \emph{B} band extinction values taken from Schlegel, Finkbeiner \& Davis (1988). Here we use $H_{0}$ = 70
km s$^{-1}$ Mpc$^{-1}$ and and $q_0$ = 0.2 for consistency with the G99
preparation of the lightcurves.

The corresponding errors in luminosity, disregarding the uncertainty
in $H_{0}$, are typically $\pm$1.6 $\times$ 10$^{41}$ erg
s$^{-1}$
and are derived from the G99 estimate of a photometric accuracy for
this band.

\subsection{Sample Refinement}
PG1226+023, also known as 3C~273, is the most
optically luminous object in this sub-sample of the PG quasars, an order
of magnitude greater in luminosity than all the other sources. This AGN is known to 
have optical, X-ray and radio jets. It is the only
source in the optically selected sample which shows definite jets,
though 7 of the 42 objects are radio-loud. The jet emission is thought
to be synchrotron and variable itself. Since we cannot separate the
 disc emission from that of the jet PG1226+023 will not be
included in this analysis.

\section{The Model}
We assume the optical emission observed in the PG sample originates from an
accretion disc.  
Adopting the approach of SC89, the disc may be divided into two
distinct regions: the optically thick outer region emitting blackbody
contributions into the \emph{B} band, and the inner region of the disc, which
does not significantly contribute to the \emph{B} band emission. \\
We assume that the optical variability is caused by an instability on the
thermal time-scale, which varies the position of the inner radius of the optical emission
region. We assume that the two-folding time-scales we measure in the \emph{B} band
lightcurves of the PG sample correspond to the thermal time-scale at the radius
$R_{\rm cut}$ where the luminosity in the \emph{B} band has decreased by 50 per
cent from its maximum value. The value of the radius $R_{\rm cut}$ for a given
luminosity depends upon the disc model chosen, accretion rate and black hole mass.\\
Comparing $R_{\rm cut}$ for each model with the thermal time-scale, $\tau_{\rm th}$, gives a value of
$\alpha$ from the following equation
\begin{equation}
                 \tau_{\rm th} = \frac{1}{\alpha  \omega_{\rm k}}
\end{equation}
where $\omega_{\rm k}$ = $\sqrt \frac{GM}{R^{3}}$ is
the Keplerian angular velocity. 

\begin{figure}
\centering
\vspace*{7.5cm}
\leavevmode
\includegraphics{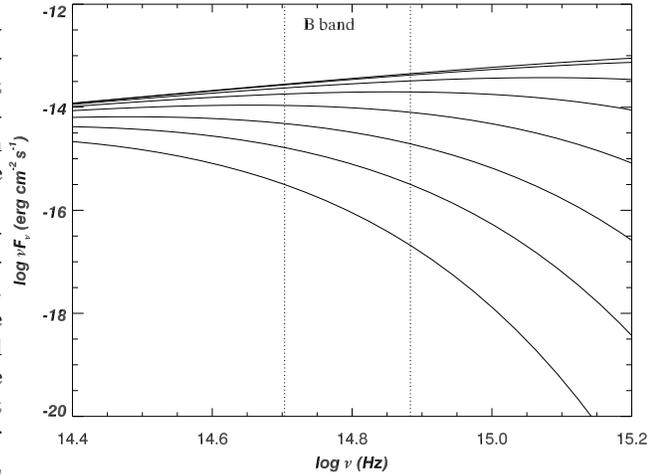}
\caption{The model blackbody spectrum for a $M = 10^8\msun$ black hole,
accreting at one tenth of the Eddington rate, with increasing radial disc size
$R_{\mathrm{cut}}$ = 3 (upper curve), 10, 29, 64, 127, 218, 345, 514 (lower curve)
$R_\mathrm{g}$. The outer disc radius is always fixed at $R_{\mathrm{outer}} =
10^{3}R_\mathrm{g}$.\newline The vertical dotted lines mark the region covered by the
\emph{B} band in the observers' frame (3600-5600$\angstrom$).}
\label{modelflux}
\end{figure}

To describe the outer optical flux emitting disc we use a standard multi-temperature blackbody $\alpha$-disc code
developed by Czerny \& Elvis (1987) and used in SC89.
The model assumes a geometrically thin, optically thick Keplerian disc
accreting steadily onto a Schwarzschild black hole. \\
We assume that the disc radiates locally as a blackbody with an emitted flux of \begin{equation}
 \sigma T_{\rm eff}^{4}~=~\frac{3GM \dot{M}}{8\pi R^{3}}\left(1-\sqrt{\frac{R_{\rm inner}}{R}}\right)
\end{equation}
where $R$ is the radial distance from the central object, $\sigma$ is the
Stefan-Boltzmann constant, $T_{\rm eff}$ is the effective temperature of the disc
at radius $R$, $G$ is the gravitational constant, $M$ is the black hole mass,
$\dot{M}$ is the steady accretion rate and $R_{\rm inner}$ is the inner radius
of
the disc (eg. Frank, King \& Raine 1992). We assume $R_{\rm inner}=3R_{\rm g}$, where $R_{\rm g} \equiv 2GM$/$c^{2}$.
Inside the radius $R_{\rm cut}$ we do not make any further assumptions
about the nature and spectral properties of the disc. We only assume that the optical flux
from that region is negligible. Fig. $\ref{modelflux}$ shows different spectra depending on
the choice of $R_{\rm cut}$.
Note that the
maximum contribution to the flux at a given frequency $\nu$ comes from
the disc region with the temperature $kT_{\rm eff}$ = $1.65 h \nu$, which is
located at the distance $R_{\nu} = 0.77\times 10^{24} \nu
^{-4/3}(M/\msun)^{-1/3}(L/L_{\rm Edd})^{1/3} R_{\rm g}$ (SC89), where
$L_{\rm Edd}\sim1.3\times 10^{38} (M/\msun)~~\ergsec$ is the Eddington
luminosity. Hence for a source with a black hole mass of
$10^{8}\msun$ emitting at the Eddington luminosity the location of
the \emph{B} band is $R_{\nu} \sim 21-38 ~R_{\rm g}$, within our chosen range.

The shape of the spectrum depends upon  \\
1) $M$, the black hole mass  \\
2) $\dot{M}$, the accretion rate  \\
3) the extent of the outer disc contributing to the optical flux, which depends
   on the variable $R_{\rm cut}$ since $R_{\rm outer}$ is fixed at
10$^{3}$ R$_{\rm g}$\\  
4) $i$, the disc inclination. We assume an inclination angle of $\cos{i}$ = 0.75 (41$^{\circ}$) since
optical spectroscopy has shown that all the sample objects have broad emission lines.

\begin{figure}
\centering
\vspace*{7.5cm}
\leavevmode
\includegraphics{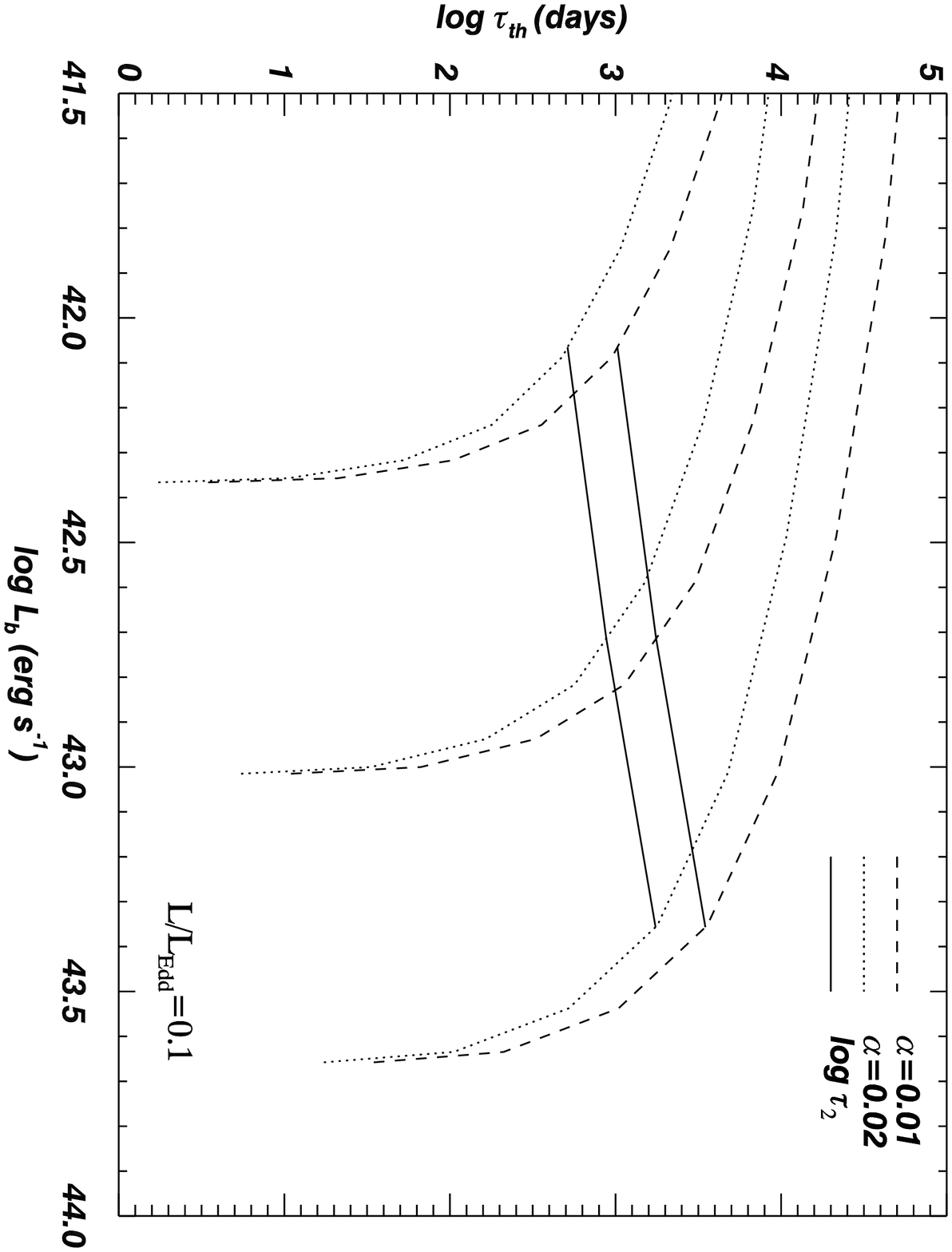}
\caption{This figure shows the blackbody disc models' \emph{B} band luminosity $vs.$ thermal time-scale. \newline
This is the case of $L/L_{\mathrm{Edd}}$ = 0.1 and black hole masses
$10^{7},10^{7.5}$ and $10^{8} \msun$, with the smallest mass producing the lowest
luminosity. \newline
For each mass the model is calculated for 8 steps in $R_{\mathrm{cut}}$ from the last stable orbit
(3$R_\mathrm{g}$) out to 512$R_\mathrm{g}$. This has been done for two values of the viscosity parameter;
$\alpha$ = 0.01 (dashed lines) and $\alpha$ = 0.02 (dotted
lines). The solid lines join the points at which the luminosity has decreased 
by 50 per cent from its maximum value in each model ($\tau_\mathrm{2}$).}
\label{modeleg}
\end{figure}

We also consider the effects of opacity in a modified blackbody model
in which electron scattering in the disc is taken into
account (Czerny \& Elvis 1987; Haardt \& Maraschi 1993).
Electron scattering becomes important in the disc at high
temperatures, so we expect the effects of opacity to be
visible at ultraviolet energies and above, but should not significantly modify the
continuum emission at optical wavelengths.

To enable comparison with the data the luminosities of the model
spectra were convolved with the \emph{B} filter which had been applied during
observation of the PG sample. The integrated luminosity over the Johnson
\emph{B} band, $L_{B}$, is calculated and directly compared with the observed
\emph{B} luminosities. Two-folding time-scales for the models
are readily obtained from the $\log L~vs.~\log \tau_{\rm th}$ plots by
measuring the time taken for the luminosity to drop from its maximum
to half that value (Fig. $\ref{modeleg}$). Each value of $\alpha$ for a set of
models has a characteristic time-scale (Equation 6). Finding
sets of models which encompass the data sample allows the
determination of the range of the viscosity parameter for that data
set. 

\section{Results}
\subsection{Variability}
\begin{figure}
\centering
\vspace*{7.5cm}
\leavevmode
\includegraphics{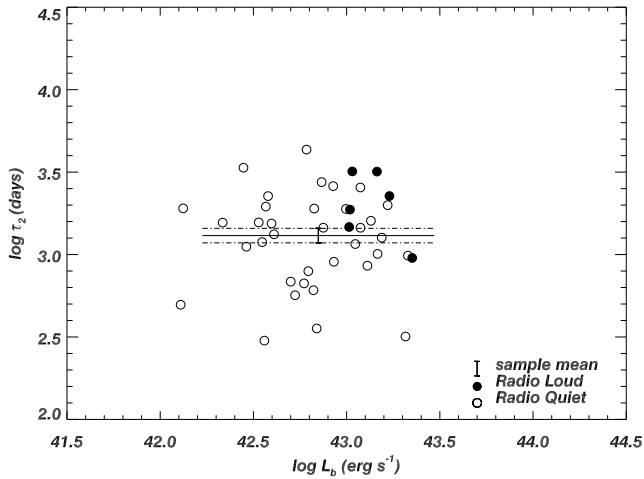}
\caption{The two-folding time-scales for each source versus
luminosity in the \emph{B} band (log scale). The mean time-scale and standard
deviation of the mean are represented by the solid line and dot-dashed lines
respectively, with the vertical error bar at the point of the mean luminosity. The
uncertainty in the value of $H_{0}$ introduces uncertainties in the
luminosities, so we have not calculated the error on the mean luminosity here. The length of the luminosity error bar includes 95 per cent of the data (2$\sigma$).}
\label{meanlog}
\end{figure}

\begin{figure}
\centering
\vspace*{7.5cm}
\leavevmode
\includegraphics{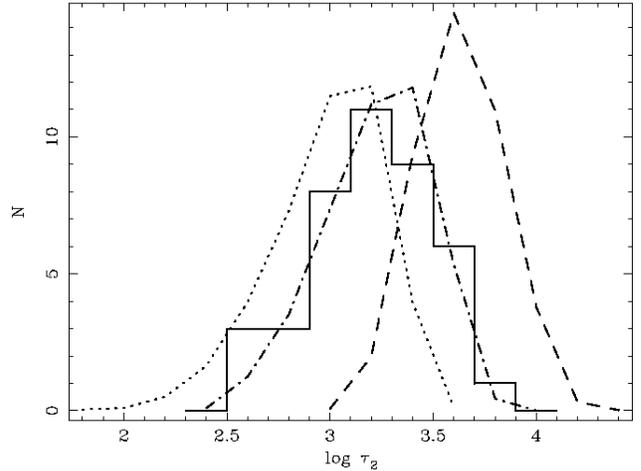}
\caption{Comparison of the observed distribution of two-folding time-scales for 41
PG sample AGN (binsize $\Delta \log \tau_\mathrm{2}=0.2$, $N$ counts
per bin) with the distributions predicted from random realisations of the
broken power law power
spectrum (with break frequency $\nu_{\mathrm{bk}}$) which probably describes
the red-noise optical variability in
AGN (10000 lightcurves simulated, $N$ rescaled to 41 counts over
whole distribution). \newline Solid histogram: observed distribution, dotted line:
$\nu_{\mathrm{bk}}=10^{-7}$~Hz, dot-dashed line: $\nu_{\rm
bk}=2\times 10^{-8}$~Hz, dashed line: $\nu_{\mathrm{bk}}=4\times 10^{-9}$~Hz.}  
\label{numcts}
\end{figure}

The lightcurves for this dataset (G99) show a typical intrinsic variability
amplitude of $\sigma_{B}$ = 0.14 mag with intrinsic rms amplitudes
of 5$\%<\sigma_{B}/B<$34$\%$ over the seven year period.
The variability time-scales lie approximately between 300 and 4500
days.
The distribution of two-folding time-scales is shown in Figs.
$\ref{meanlog}$ and $\ref{numcts}$ and the individual time-scales and luminosities for each AGN
are given in Table 2. The mean two-folding time-scale for the sample is 1566.4 days and the corresponding standard deviation of the
mean is 143.6 days (in log space, mean log $\tau_{\rm 2} = 3.12\pm0.04$). 

\begin{center} 
\begin{table}
\scriptsize
\caption{Individual minimum two-folding time-scales and luminosities for AGN in
this PG subsample. The luminosity quoted is the median \emph{B} band luminosity
in the rest-frame of the source. Errors in the luminosities stem from errors in the observed
magnitudes and hence the flux. Errors due to the uncertainty in $H_{0}$ are not included here.} 
\begin{tabular}{ccccccc}  \hline 
\scriptsize
Quasar&other names&z&RL&L$_{4400\ang}\times$10$^{42}$ &$\tau_{\rm 2}$ \\ 
 & & & & (erg s$^{-1}$) & (days) \\ \hline
PG0026+129 & &0.142& &11.9$\pm$0.22&1454 \\ 
PG0052+251& &0.155& &20.7$\pm$0.38&318 \\ 
PG0804+762& &0.100& &14.7$\pm$0.27&1009 \\  
PG0838+770& &0.131& &1.33$\pm$0.02&1907 \\ 
PG0844+349&Ton 951&0.064& &5.29$\pm$0.10&567 \\  
PG0923+201&Ton 1057&0.190& &11.1$\pm$0.20&1158\\  
PG0953+415&K 348-7&0.239& &15.4$\pm$0.29&1266  \\  
PG1001+054& &0.161& &3.62$\pm$0.07&301 \\  
PG1012+008& &0.185& &16.6$\pm$0.31&1994\\  
PG1048+342& &0.167& &1.29$\pm$0.02&496 \\   
PG1100+772&3C 249.1&0.313&Y&14.5$\pm$0.27&3186 \\    
PG1114+445& &0.144& &6.70$\pm$0.12&1901 \\  
PG1115+407& &0.154& &3.80$\pm$0.07&2262 \\   
PG1121+423& &0.234& &6.23$\pm$0.11&793 \\  
PG1151+117& &0.176& &6.64$\pm$0.12&608 \\  
PG1202+281&GQ Com&0.165& &5.02$\pm$0.09&685\\   
PG1211+143& &0.085& &13.5$\pm$0.25&1602\\  
PG1229+204&Ton 1542&0.064& &3.52$\pm$0.06&1189\\  
PG1307+086& &0.155& &8.54$\pm$0.16&904 \\   
PG1309+355&Ton 1565&0.184&Y &10.4$\pm$0.19&1874 \\  
PG1322+659& &0.168& &6.10$\pm$0.11&4329\\  
PG1351+640& &0.087& &9.92$\pm$0.18&1891 \\  
PG1354+213& &0.300& &3.69$\pm$0.07&1953\\   
PG1402+261&Ton 182&0.164& &8.47$\pm$0.16&2600\\  
PG1404+226& &0.098& &2.16$\pm$0.04&1562\\  
PG1411+442&PB 1732&0.089& &7.35$\pm$0.14&2750 \\  
PG1415+451& &0.114& &2.90$\pm$0.54&1116\\  
PG1426+015&Mkn 1383&0.086& &4.08$\pm$0.07&1327\\  
PG1427+480& &0.221& &3.95$\pm$0.07&1542\\  
PG1444+407& &0.267& &2.80$\pm$0.05&3362\\   
PG1512+370&4C 37.43&0.371&Y &10.7$\pm$0.20&3194\\    
PG1519+226& &0.137& &3.38$\pm$0.06&1566\\  
PG1545+210&3C 323.1&0.266&Y &10.3$\pm$0.19&1470\\   
PG1613+658&Mkn 876&0.129& &12.9$\pm$0.24&856\\  
PG1617+175&Mkn 877&0.114& &6.91$\pm$0.13&356\\  
PG1626+554& &0.133& &5.91$\pm$0.11&668 \\   
PG1700+518& &0.292& &21.3$\pm$0.40&985\\  
PG1704+608&3C 351&0.371&Y &22.5$\pm$0.41&954\\   
PG2130+099&II Zw 136&0.061& &7.50$\pm$0.14&1454\\  
PG2233+134& &0.325& &11.8$\pm$0.22&2551\\   
PG2251+113&PKS 2251+11 &0.323&Y &17.00$\pm$0.31&2264\\  
\end{tabular}
\end{table}
\end{center}

\subsection{Distribution of two-folding time-scales}
The optical variability of AGN 
is consistent with a red-noise process,
so that each lightcurve is a realisation of an underlying process
described by a power law power spectrum with a slope and
amplitude determined by the
mechanism which causes the variability. In principle, it is
possible to derive the power-spectral parameters from the
model for variability, and compare these with observations, however for
simplicity and ease of interpretation we consider only the two-folding
time-scale, which is a property of the power-spectral shape. Due to the
stochastic nature of red-noise processes, lightcurves obtained at
different times will look different, and show different observed two-folding
time-scales.
Ideally, we need to measure the `true' average two-folding
time-scale by averaging the two-folding time-scale over many
7-year lightcurves from the same source. As this is clearly not possible,
we must take account of the fact that the
two-folding time-scale measured for a given source is not the same as the
average two-folding time-scale, and considerable scatter is
introduced into the distribution of observed two-folding time-scales for
sources which have identical variability properties (i.e. identical      
power-spectral shapes). \\
We can test whether the observed distribution of two-folding time-scales is
consistent with a single power-spectral shape and amplitude (and a
corresponding single average two-folding time-scale), by comparing the
distribution we observe
with that determined from simulated red-noise lightcurves generated from
plausible power-spectral parameters. 
We therefore simulated random realisations of red-noise lightcurves using
the method of Timmer \& K\"{o}nig (1995). To date, only the X-ray
power-spectral shapes of AGN have been well constrained \citep{phil1}, showing fairly steep
($1/\nu^{2}$) power-law slopes at high frequencies, breaking to flatter
$1/\nu$ slopes below some break frequency. The optical
power-spectral
shape of the AGN in our sample is not well constrained, but correlated
long-time-scale optical/X-ray variations in the luminous Seyfert galaxy
NGC~5548 suggest that although the optical power spectrum of luminous
AGN is steeper than the X-ray power spectrum at high frequencies, the
shapes are similar at low frequencies \citep{phil2}. Therefore we
assume high-frequency power-spectral slopes of $1/\nu^{2.5}$, breaking to
$1/\nu$ below some break frequency, $\nu_{\rm bk}$, which we assume is
common to both X-ray and optical power spectra. Breaks in the X-ray
power spectra of AGN appear to scale linearly with black hole mass
\citep{phil1}, so we first chose a break
frequency of $2\times10^{-8}$~Hz, corresponding to a black hole mass of a
few $\times~10^{8}$~M$_{\odot}$. 
We chose a power-spectral normalisation which corresponds to the typical observed  
$\sim10$ per cent variability amplitude in the sample. \\
We simulated 10000
lightcurves of 7~year duration, and measured the resulting distribution of
two-folding time-scale $\tau_{\rm 2}$, which is plotted for comparison with the
observed distribution of $\tau_{\rm 2}$ in Fig. $\ref{numcts}$. We also show
the distributions corresponding to break frequencies of $10^{-7}$~Hz and
$4\times10^{-9}$~Hz, i.e. a possible factor $\sim25$ range in black hole
mass. The observed distribution of $\tau_{\rm 2}$ is consistent with
that expected from random realisations of a single power spectral shape
expected from few $\times~10^{8}$~M$_{\odot}$ black holes. So the
distribution of black hole masses is likely to be narrow ($<$~decade) and the
variability in the lightcurves of all the AGN can be characterised by a single identical two-folding
time-scale, as
expected from a homogeneous sample of AGN such as that presented here. This result is not
sensitive to the slope above the break. From
Equation 6 the sample must also be consistent with a single value of the
quantity $R_{\rm cut}^{3/2}/\alpha$. Keeping the Eddington ratio $L/L_{\rm Edd}$ fixed, there
is a single value of the
viscosity parameter $\alpha$ determined by the average $\tau_{\rm 2}$ of the
sample. 
      
For $0.01~\le~L/L_{\rm Edd}~\le~1.0$ the data are
best fitted by the set of models outlined in Table 2.
\begin{figure*}
\centering
\vspace*{7.5cm}
\leavevmode
\includegraphics{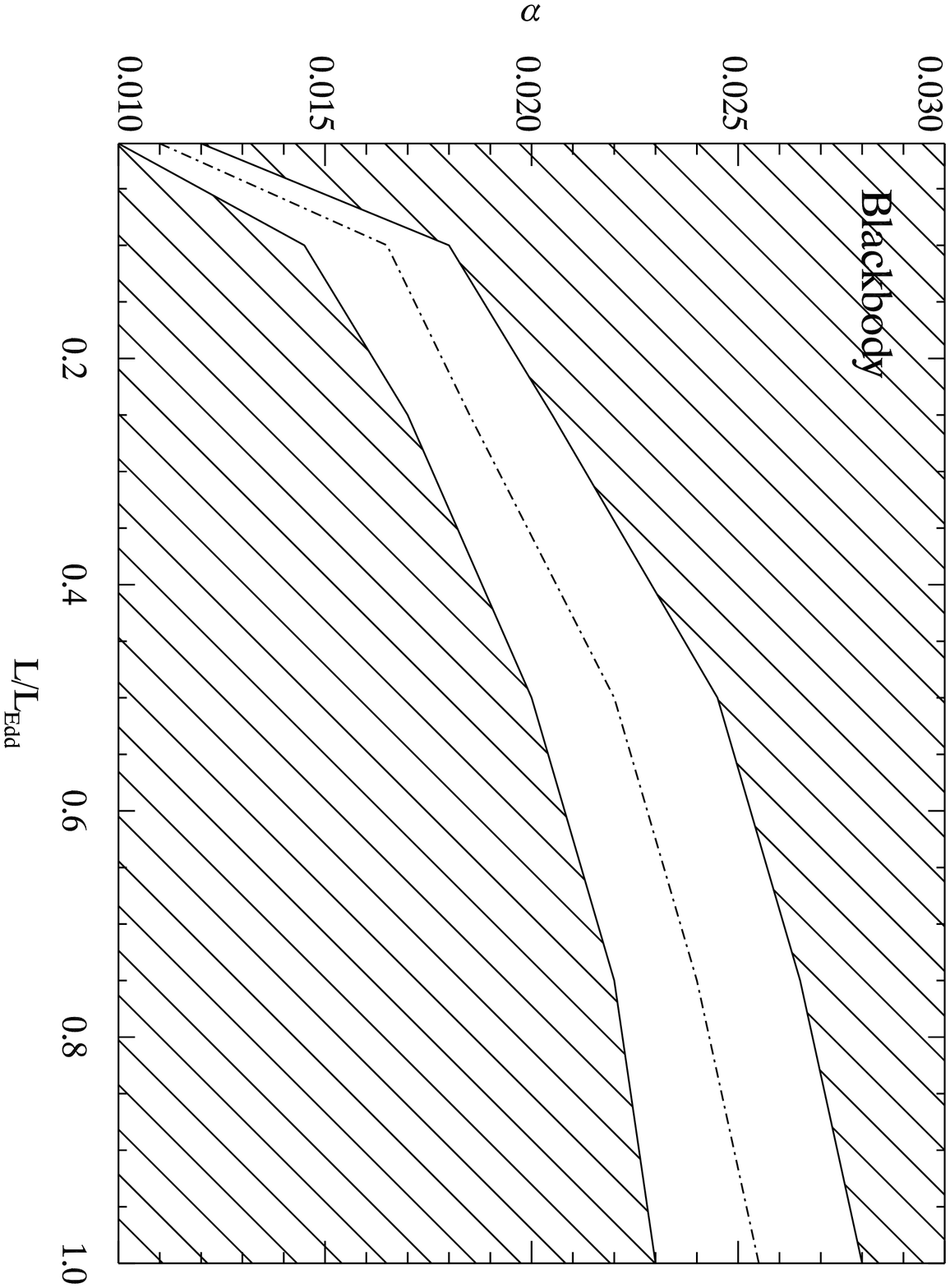}
\hspace*{9cm}
\includegraphics{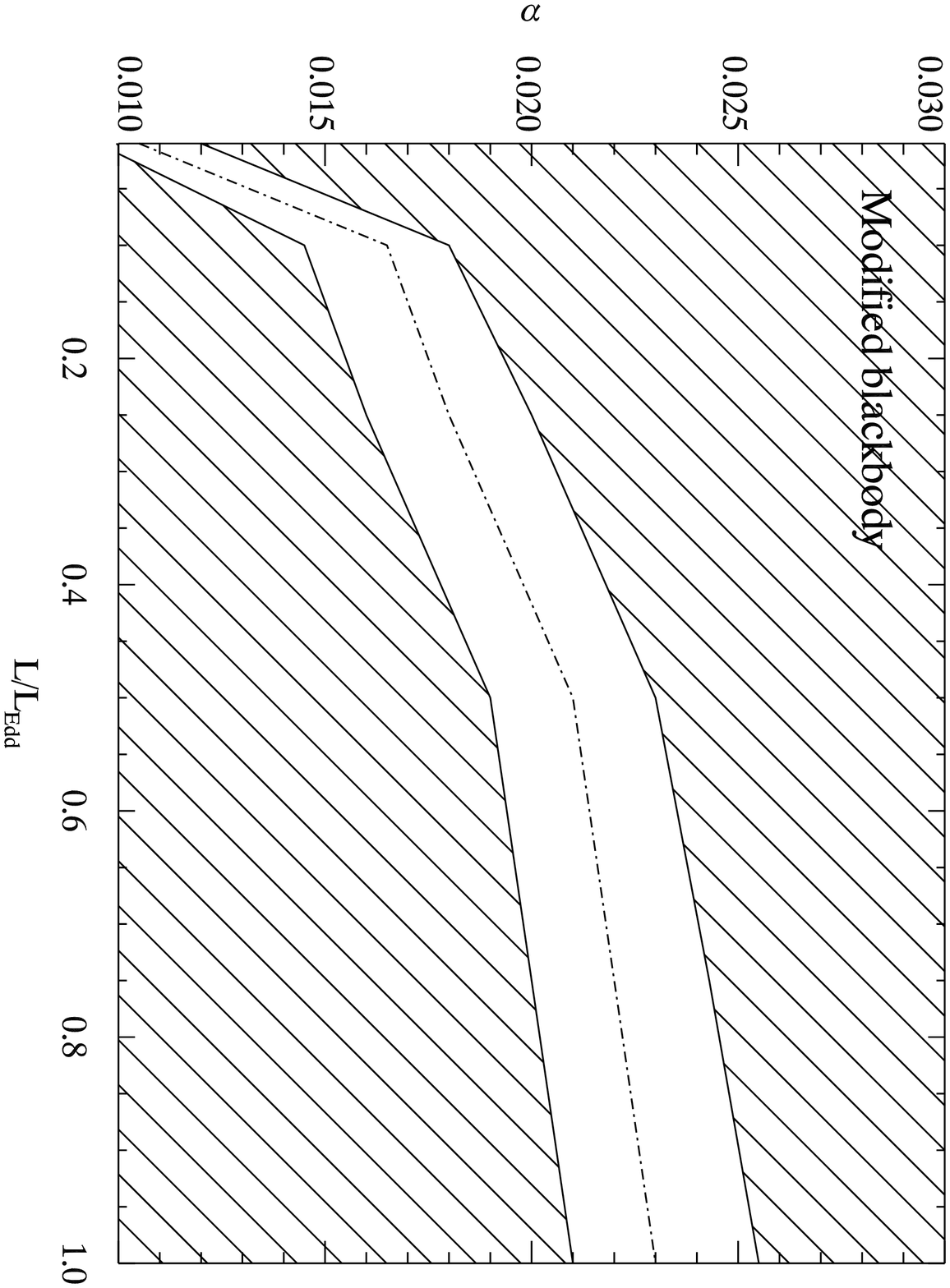}
\caption{This figure shows the allowed range of $\alpha$ values for both the
blackbody models (left) and the modified blackbody models (right). The permitted values of $\alpha$ lie within the solid
lines, the best fitting values lying along the dot-dashed line. The shaded
areas are regions in which a given viscosity parameter, $\alpha$, and Eddington
 ratio, $L/L_{\rm Edd}$, cannot occur simultaneously in the model.}  
\label{alpha}
\end{figure*}
Fig. $\ref{alpha}$ shows the allowed values of $\alpha$ over a range of the
Eddington ratio for both the disc models we apply.
We note that
the difference in viscosity between blackbody and modified
blackbody models is small over this wavelength range.
\begin{center}
\begin{table}
\caption{Model parameters for the PG data, with corresponding mass and
accretion rate ranges of $10^{7}<M<10^{8.5}$ ($\msun$) and
$0.02<\dot{M}<0.70$ ($\msun$ year$^{-1}$).}
\begin{tabular}{@{}lcccc} \hline
Model&$L/L_{\rm
Edd}$&$\alpha_{\rm min}$&$\alpha_{\rm mean}$&$\alpha_{\rm max}$ \\ \hline
disc blackbody& 0.01& 0.010 & 0.011  &0.012\\
disc blackbody& 0.1 & 0.015 & 0.017 & 0.018\\
disc blackbody& 0.25& 0.017 & 0.019 & 0.021 \\
disc blackbody& 0.5 &0.020  &0.022 & 0.025 \\
disc blackbody& 0.75 &0.022 & 0.024  &0.027\\
disc blackbody& 1.0 & 0.023  &0.026  &0.028 \\
modified blackbody & 0.01& 0.010  &0.011  &0.012\\
modified blackbody & 0.1& 0.015  &0.017  &0.018  \\  
modified blackbody & 0.25& 0.016 & 0.018 & 0.020 \\
modified blackbody & 0.5&0.019  &0.021  &0.023 \\
modified blackbody & 0.75&0.020  &0.022  &0.024   \\
modified blackbody & 1.0 & 0.021 & 0.023 &0.026\\  
\end{tabular}  
\end{table}
\end{center}
 
\section{Discussion} 
Since direct observations cannot yet resolve the inner accretion disc in AGN, numerical simulations of accretion flows have existed for several decades and
are moving in the direction of global 3-D models. Our results fall within the
range $0.005 \le \alpha \le 0.6$ predicted by a variety of recent numerical
simulations of MHD turbulent discs. However, we obtain much tighter constraints
towards the lower end of this broad range which could eliminate some models. For
example, Armitage (1998) writes that $\alpha\sim0.1$ in global and local
cylindrical disc simulations with net vertical field, whereas $\alpha\sim0.01$
when the net vertical field is zero. MHD simulations have also shown that an
alternative $\alpha$-prescription in which stress is proportional only to the
magnetic pressure may be a better diagnostic of the
viscosity mechanism, since using the Shakura Sunyaev (1973) $\alpha$-prescription a range of $\alpha$ values can be obtained for the
same gas pressure \citep{HB95}.  

The only previous observational viscosity estimates for AGN discs were those
calculated by SC89 for 8 AGN and 4 blazars
at 1060$\angstrom~$, 3 AGN at 1395$\angstrom~$ and 12 AGN at
1740$\angstrom~$, with the prediction that $\alpha_{\rm min} =$ 0.001
and $\alpha_{\rm max}\sim$ 0.1. Our results allow a much tighter estimate of
the viscosity range and we find a value around 0.02. 
Both estimates have used the same disc
model, but at different wavelengths; we chose an optical sample
rather than the UV data used by SC89 since
the optical bands lie further from the peak in the power law region of
the disc spectrum. SC89 assume an
inclination angle of 0$^{\circ}$ and include blazars in their sample.
They discuss the likelihood that the interpretation of blazar
variability as thermal changes in the accretion disc may not be valid
since synchrotron radiation is known to dominate in many of these
objects. They also note that the sampling of 5 of their sample objects
was not designed for detecting the time-scales of interest when
estimating disc viscosities.
O'Brien $\etal$ (1988) examined IUE data of several AGN
and found that most had UV two-folding time-scales of weeks to years with no shorter time-scale variability. The sampling rate for this
data set was poor. Consequently, it would not be possible to estimate $\alpha$
for their sample. 
The PG subsample used here is larger and
has a higher and more even sampling than any sample used in previous
studies of this kind.

We find this sample of 41 AGN all have similar ($<$ decade) masses and therefore can all be
represented by a single two-folding time-scale. Hence we find a very narrow
range of $\alpha$ and/or $L/L_{\rm Edd}$. 
This seems to suggest that we are observing all these AGN in a single 'state'. 
If this is true, it may be that all observed AGN are in the
outburst state and those in the quiescent state are either too faint
to have yet been detected or appear as normal galaxies
(Siemiginowska \& Elvis 1997; Burderi, King \& Szuszkiewicz 1998; Hatziminaoglou, Siemiginowska
\& Elvis 2001). If the former is true, this would then
imply that the total number of AGN in existence is far higher than
current estimates based on the number of sources we see. However, a single
value of $\alpha$ throughout the thermal limit cycle has been predicted in
some numerical simulations of MHD turbulence in AGN accretion disks \citep{MQ},
though the thermal ionisation instability is unlikely to be a source of large
amplitude variability in AGN.\\
Reverberation mapping measurements
exist for 16 of these sources \citep{Kasp} ranging from $2.16\times 10^7$ to
$4.7\times 10^8 \msun$ ($M_{\rm average}=1.51\times10^{8}\msun$) with typical
uncertainties of about 40 per cent.
Black hole masses we find through the disc modelling method are dependent
upon the ratio of AGN luminosity to the Eddington Limit, $L/L_{\rm Edd}$. For
$L/L_{\rm Edd}=0.01$, $M=10^{8}-10^{8.5}\msun$ as suggested by comparing the
observed distribution of two-folding time-scales with distributions predicted
from random realisations of the broken power law power spectra. The
reverberation mapping results are compatible with all the Eddington ratios used here.

We now summarise the main assumptions and note some caveats on our
determination of $\alpha$.
The accretion disc models we use are simple multi-temperature blackbody
spectra (Czerny \&
Elvis 1987), and we assume that the disc is the source of the \emph{B} Band
luminosity. We also assume that long term variability in this band occurs on
the thermal time-scale at a radius corresponding to a 50 per cent change in
\emph{B} Band luminosity. The limits on $\alpha$ are derived only for the sample
as a whole since the variability time-scale and mass of each individual object is
not precisely known. The results are dependent on the validity of the
assumptions stated above and require a single viscosity value throughout the
outer disc. In practice, viscosity is likely to be both radially and time
dependent and variations in luminosity may occur on various time-scales and
radii. For example, the $\alpha$ value we derive here would be a lower limit on
the true value if the observed variability time-scales are not equal to but are
longer than the thermal time-scale. 

Blackbody and modified blackbody models gave very similar (and, at
accretion rates one tenth of Eddington or lower, the same) values of $\alpha$.
The small difference between
the blackbody models and the addition of opacity effects to these models
shows that in the optical \emph{B} band, away from the peak of the emitted flux,
electron scattering has only a small effect on the spectrum as expected. \\
The disc models do not include warps, clumpiness or flares. 
Contributions to the
variability from occurrences of flares is, however, likely to be on time-scales shorter even than the sampling interval of the
lightcurves.\\
G99 estimate pollution by the host galaxies to be approximately 5 per cent
based on observations with HST \citep{HST}. The emission line contribution to
the optical flux is estimated in the same paper at 5-10 per cent. The models applied
here are for AGN continuum flux only and we have not reduced the observed
flux to account for these since a 10 per cent luminosity increase will not
greatly affect the viscosity estimates.
We also note that our luminosity estimates are affected by the choice of
$H_{0}$ which, if incorrect, would introduce an error into
our results. \\
These models do not incorporate a corona, transition
region, self-gravitation or irradiation of the disc material. Coronae
are thought to be the source of hard X-rays but could alter the shape
of the optical spectrum as we may see reprocessing and/or scattering,
and this may be important in some AGN. The role of the corona in
forming the continuum emission is discussed in Kurpiewski,
Kuraszkiewicz \& Czerny (1997) and a corona could be incorporated in future
determinations of the $\alpha$-parameter.  

There are currently no conclusive observational or
theoretical bounds on the $\alpha$-parameter. We have carried out this study in
the hope of providing another clue to the nature of viscosity in AGN.

\section{conclusions}
We assume that optical variability of AGN observed on time-scales of
months to years is caused by local instabilities occurring on the
thermal time-scale in the inner accretion disc. The disc is modelled 
using multi-temperature blackbody models.  
In this case we constrain the viscosity to  
\begin{center}
      0.01 $\le \alpha \le$ 0.03 for $0.01 \le L/L_{\rm Edd} \le 1.0$. \\
\end{center}      
We show this sample is consistent with a single $\alpha$ value at a given
Eddington ratio, determined by the
average characteristic time-scale of all 41 AGN. The range of $\alpha$ values
we find lies within the range predicted by current numerical simulations of MHD
turbulent discs. The mass range we obtain for this sample is consistent with
the reverberation mapping measurements made for 16 of these PG quasars.
However, we do stress that this is a simplified model of an
accretion disc, and merely a first step in observational viscosity estimates
for AGN.    

\section{Acknowledgments}
This work was partially supported by the NASA programs NAS8-39073 and
Chandra Award Number GO1-2117B issued by the Chandra X-ray Observatory
Center, which is operated by the Smithsonian Astrophysical Observatory for
and on behalf of NASA under contract NAS8-39073.
RLCS acknowledges support from a PPARC studentship. 
This research has made use of the NASA/IPAC Extragalactic Database (NED) which is operated by the Jet Propulsion Laboratory, California Institute of Technology, under contract with NASA.

\bsp

\label{lastpage}

\end{document}